\documentclass[reqno]{amsart}
\usepackage{graphicx,amsmath}
\usepackage{float}
\usepackage[utf8]{inputenc}
\usepackage{color}
\usepackage{cite}
\usepackage{fancyhdr}
\usepackage{url}

\begin{document}
\pagestyle{plain}
\title{Analytical and numerical modelling of ballistic heat conduction observed in heat pulse experiments}

\author{G. Balassa$^2$, P. Rogolino$^4$, Á. Rieth$^{1}$, R. Kovács$^{123}$}

\address{
$^1$Department of Energy Engineering, Faculty of Mechanical Engineering, BME, Budapest, Hungary
$^2$Department of Theoretical Physics, Wigner Research Centre for Physics,
Institute for Particle and Nuclear Physics, Budapest, Hungary
$^3$Montavid Thermodynamic Research Group, Budapest, Hungary,
$^4$Department of Mathematics and Computer Sciences, Physical Sciences and Earth Sciences, University of Messina, Messina, Italy
}

\date{\today}

\begin{abstract}
Among the three heat conduction modes, the ballistic propagation is the most difficult to model. In the present paper, we discuss its physical interpretations and showing different alternatives to its modelling. We highlight two of them: a thermo-mechanical model in which the generalized heat equation - the so-called Maxwell-Cattaneo-Vernotte equation - is coupled to thermal expansion. At the same time, the other one uses internal variables. For the first one, we developed a numerical solution and tested on the heat pulse experiment performed by McNelly et al.~on NaF samples. For the second one, we found an analytical solution that emphasizes the role of boundary conditions. This analytical method is used to validate the earlier developed numerical code.
\end{abstract}
\maketitle

\section{Introduction}
The experimental observations frequently precede the theories, which thus must be generalized or extended in some way to explain and model the observed phenomena. One long-lasting task of the physicists and engineers is to find the proper material models, including state equations and other constitutive relations. In the following, we are focusing on the heat conduction theories and their experimental motivations.

One counterexample for the above, the theoretical prediction for the existence of the so-called second sound - the damped wave propagation form of heat - preceded the experimental observation \cite{JosPre89, JosPre90a}. That prediction found by Tisza and Landau, earlier of the XX.~century \cite{Tisza38, Lan41}, and the first successful experiment is performed by Peshkov in 1944 \cite{Pesh44}.

However, this was not the case for the ballistic heat conduction. McNelly et al.~\cite{JacWalMcN70, JacWal71, McN74t} were among the first ones who experimentally detected a heatwave with the speed of sound. This is one characteristic property of that propagation mode. Although dozens of successful measurements are performed, and that particular phenomenon is observed in various situations, its theoretical background is still under development, since about 50 years.

During that era, several approaches are developed, explicitly dedicated to model the ballistic heat conduction \cite{DreStr93a, Ma13a, Ma13a1, KovVan15, CimFri96, JouEtal10b, AlvJou07KN}, which are discussed briefly in the following section.
In the meantime, the second law of thermodynamics emerged to be as an effective tool to find the proper background with generalizing the classical Fourier's law. In regard to the ballistic conduction, we have to mention three branches that are both historically and scientifically appeared to be essential.

The first one is the so-called Extended Irreversible Thermodynamics (EIT), using dissipative fluxes (such as the heat flux) as new non-equilibrium state variables \cite{Cimmelli09nl, Cimm09diff, JouEtal10b, RogCimm19, JouVasLeb88ext}. That framework is built-on the balance laws, and exploiting the second law to derive the constitutive equations. This is similar to the approach of the Non-Equilibrium Thermodynamics with Internal Variables (NET-IV) \cite{VanFul12, BerVan17b, Verhas97, Verhas96, RogEtal17,  JozsKov20b, Nyiri89, Gyar77a, Gyarmati70b}. In fact, NET-IV goes further in the generalization, it says that one must not attach any direct physical meaning to the new state variable, but only its tensorial order. It allows a more general, a more flexible background \cite{KovEtal18rg, Kov18rg}, allowing more possibilities of coupling between different field variables. That flexibility disappears from the framework of Rational Extended Thermodynamics (RET), built-on the kinetic theory \cite{MulRug98}. Here, an infinite hierarchy is built in the form of balance laws based on the Boltzmann equation. Then the second law emerges in the closure procedures together with Galilean invariance \cite{MulRug98, RugSug15}. This results in more cumbersome calculations, but favourable in many physical situations \cite{PavicEtal16, AriEtal12c}. As an outcome, the validity range of the resulting model is restricted for specific states, thus losing some things from its generality. On the other side, it remains as a predictive theory and offers an extra knowledge about the material coefficients instead of finding them by fitting to experiments (some of them still must be fitted, but not all of them as in EIT and NET-IV).

All these approaches have their advantages and disadvantages, but this paper focuses on a different aspect. The generalized heat equations - due to the extended relationship between the heat flux and the temperature gradient - require more care about the definitions of initial and boundary conditions. In many situations, the initial conditions are homogeneous and related to a steady-state, as we also consider in the following.

However, the treatment of boundary conditions is not that simple. For instance, and this is what we concentrate on, in a heat pulse experiment, the heat flux is prescribed in time on the boundary. While the exact shape of the heat pulse can be approximated with a function more suitable for either analytical or numerical calculations, it cannot be defined using the temperature (or its gradient, accordingly). Since the parallel prescription of both the temperature and the heat flux on the boundary is still an open mathematical question, it is favourable to avoid using the temperature as a variable. This itself is not the complete picture.

Utilizing any of these approaches, one arrives at a system of partial differential equations (PDE), with three field variables: temperature (scalar), heat flux (vector), 'thermal pressure' (second-order tensor). These models differ in their derivation method, and hence in the interpretation of the phenomena \cite{KovVan18}. It also limits their validity and forms the previously mentioned properties, discussed in the next section.

\section{Physical interpretations of ballistic conduction}
So far, EIT, NET-IV and RET models appeared in the discussion. From now on, we consider the approach of EIT to be as a particular case of NET-IV and not described separately. Instead, we leave the space for other models as well.
Thus first, we present the basic ideas behind each approach and the interpretation of ballistic conduction. Then, we continue with the results of McNelly et al.~\cite{McN74t, McNEta70a}, together with the experimental setup. After shortly introducing the background of each approach aimed to model ballistic heat conduction, we focus on the analytical solutions for the NET-IV model. Since there are compatibility conditions between EIT, NET-IV and RET approach \cite{KovEtal18rg}; the presented analytical solution works for the other models as well by keeping these conditions in line of sight.

\subsection{Rational Extended Thermodynamics}
The approach of RET is considering the kinetic theory rigorously, and hence inheriting a description based on a distribution function. Its space-time evolution is prescribed by the Boltzmann equation, which is often too difficult to solve directly. Instead, applying the momentum series expansion yields an infinite series of PDE, consisting of increasing tensorial order quantities.

Particularly for heat conduction problems, the phonon hydrodynamics is utilized by apriori prescribing their interactions. Unfortunately, such approach restricts the validity of the model to rarefied situations, for instance, low temperature problems. At the end, it yields in one spatial dimension \cite{MulRug98}
\begin{equation}
\frac{\partial u_{\langle m \rangle}}{\partial t} + \frac{m^2}{4m^2-1}c_D\frac{\partial u_{\langle m-1 \rangle}}{\partial x}+c_D\frac{\partial u_{\langle m+1 \rangle}}{\partial x}=\left \{ \begin{array}{ll}
\displaystyle
0 \ & \ m=0 \\
-\frac{1}{\tau_R}u_{\langle 1 \rangle} \ & \ m=1 \\
- \left( \frac{1}{\tau_R}+\frac{1}{\tau_N} \right )u_{\langle m \rangle} \ & \ 2\leq m\leq M
\end{array} \right. \label{phhysys}
\end{equation}
in which $c_D$ is the Debye speed of phonons, and $u$ denotes the corresponding momentum quantity, i.e. $u$ is the zeroth order (scalar) momentum, $u_i$ is a vectorial one, and it continues to infinity, i.e., $M\rightarrow \infty$; closed with truncation. Furthermore, $_{\langle \ \rangle}$ denotes the traceless symmetric part of a tensor, and  the subscript $m$ indicates the $m^{th}$ momentum $u_{<m>}$.
The remarkable moments - using the notations of Dreyer and Struchtrup - are the following \cite{DreStr93a}:
\begin{itemize}
  \item energy density: $e=hcu$;
  \item momentum density: $p_i=hu_i$;
  \item energy flux, which is proportional with the momentum density: $Q_i=hc^2u_i$;
  \item deviatoric part of the pressure tensor: $N_{\langle ij \rangle}=hcu_{\langle ij \rangle}$.
\end{itemize}
In eq.~\eqref{phhysys}, $\tau_R$ and $\tau_N$ are the relaxation times for resistive and normal phonon interactions \cite{MulRug98}. Only these quantities are to be fit.

Eq.~\eqref{phhysys} is tested on McNelly's experimental data, using $M=2$ \cite{DreStr93a}. This gives a characteristically good agreement; however, the propagation speeds are inappropriate since the fixed coefficients in \eqref{phhysys} mean a strict constraint: the speed of sound, i.e., the speed of the ballistic signal is recovered only with $M\rightarrow \infty$, but fairly approximated with $M=30$ \cite{MulRug98, KovVan16, KovVan18, Van16}.

\subsection{Non-Equilibrium Thermodynamics with Internal Variables}
On the contrary to the RET, it requires no assumptions about the exact mechanisms behind the phenomena. Instead, it assumes the basic balances to be valid and exploits them when solving the second law inequality \cite{VanFul12, FulEta14m1}. For details of the derivation, we refer to our earlier papers \cite{KovVan15, JozsKov20b}.

Since there is nothing said at the beginning about the mechanisms, but the entropy is defined only, the second law offers such parabolic-dissipative closing for the balance equations that are compatible with the approach of RET in a particular case, namely when the system is reduced to a hyperbolic one \cite{Van16}. NET-IV is like a parabolic envelope for the RET models, at least in the linear regime. In parallel, it allows such freedom to adjust the ballistic propagation speed by not restricting the coefficients.

The outcome of NET-IV is called ballistic-conductive (BC) model and reads as
\begin{eqnarray}
\rho c \partial_t T + \partial_x q & = & 0 \nonumber, \\
\tau_q \partial_t q +  q + \lambda \partial_x T + \kappa_{21} \partial_x Q &=& 0  \nonumber , \\
\tau_Q \partial_t Q +  Q - \kappa_{12} \partial_x q&=& 0 . \label{bceq}
\end{eqnarray}
in one spatial dimension. Here, $q$ is the heat flux, $Q$ being the current density of heat flux, $T$ stands for the temperature, $\rho$ and $ c$ are the mass density and specific heat, $\tau_q$ and $\tau_Q$ are the relaxation times corresponding to the respective fields, $\kappa_{21}$ and $\kappa_{12}$ form the antisymmetric part of the corresponding Onsager conductivity matrix, i.e. $\kappa_{21} = - \kappa_{12}$ and called 'dissipation parameter' \cite{KovVan15}. The coefficient $\lambda$ stands for the thermal conductivity. In this particular case, eq.~\eqref{bceq} is compatible with \eqref{phhysys} ($M=2$) by directly comparing the coefficients to each other.

While phonon hydrodynamics from RET says that ballistic propagation is a non-interactive process among phonons and the boundary scattering is what matters, the NET-IV based continuum models interpret it as a thermo-mechanical phenomenon due to the appearance of  the speed of sound. The $Q$ in eq.~\eqref{bceq} is kind of a 'thermal pressure'. However, this model of NET-IV is not explicitly a thermo-mechanical one since mechanics is indeed missing, only its 'thermal part' is kept.
One step forward could be made by adding the mechanical part, too, for instance, including thermal expansion as a most straightforward reversible thermo-mechanical phenomenon that can generate ballistic conduction, the elastic sound wave that coupled to the thermal field.

\subsection{Heat conduction with thermal expansion}
In the classical approach, the Duhamel-Neumann body is the simplest example of how to include and couple mechanics to the thermal field. In that case, the thermal expansion appears in the pressure tensor only. The next step is to strengthen the coupling by adding the corresponding mechanical contributions to the internal energy; that is, the mechanical motion affects the temperature field. This is a two-way (strong) coupling. However, the second sound is still absent.

In order to overcome that problem, one could imply the first generalization of the Fourier's law, called Maxwell-Cattaneo-Vernotte (MCV) equation \cite{Max1867, Vernotte58, Cattaneo58}. The second law forbids to couple the heat flux to the velocity; thus, the constitutive equations of the thermal and mechanical fields remain uncoupled, except at the level of state functions. More importantly, it turned out recently, that specific nonlinearities in the MCV equation - especially the temperature-dependent coefficients - result in the temperature dependence of mass density, i.e., thermal expansion effects \cite{KovRog20}. Consequently, it requires mechanical coupling, too. Although we are dealing with a linear system here, it is essential to investigate to understand the nonlinear behaviour of generalized heat equations.

For simplicity, let us remain in 1 spatial dimension, and represent the internal energy $e$ as \cite{FulThermo19}
\begin{align}
  e=cT+\frac{E}{2\rho} \varepsilon^2+\frac{E\alpha}{\rho}(T-T_0) \varepsilon,
\end{align}
in which $E$ is the Young's modulus, $\alpha$ is the thermal expansion coefficient, $T$ being the temperature, with $T_0$ indicating a reference one, and $\varepsilon$ is the strain.
In that case, the balance of internal energy is
\begin{align}
  \rho \partial_t e+\partial_x q = \sigma \partial_t \varepsilon,
\end{align}
where $\sigma$ is the stress, and the mechanical work is modelled as a volumetric heat source. Furthermore,
\begin{align}
  \sigma=E \varepsilon - E \alpha (T-T_0)
\end{align}
includes the thermal expansion effect as well. Finally, attaching the momentum balance, the kinematic relation, and the MCV equation, respectively,
\begin{align}
  \rho \partial_t v-\partial_x \sigma&=0, \\
  \partial_t \varepsilon &= \partial_x v, \\
  \tau_q \partial_t q + q &=-\lambda \partial_x T,
\end{align}
where only small deformation regime is considered with constant mass density $\rho$. That seems to be the simplest thermo-mechanical system which consists both the first and second sound effects, let us refer to it as MCV-TE model. Now it makes the physical background clearer than the internal variable approach, since every quantity possess a specific meaning, even in the classical case.

We mention here three further remarkable examples for thermo-mechanical formulations.
\begin{itemize}
  \item The recent work of B. Tóth \cite{TothB18} draws our attention to the variational formulation of thermoelasticity, also including second sound. There is a significant effort in this work to put the background on a solid mathematical basis. Moreover, the possible formulations of boundary conditions - due to the variational principle - is analyzed in detail.
  \item The work of Frischmuth and Cimmelli is based on the concept of non-equilibrium temperature \cite{FriCim95, FriCim96, FriCim98}, which is connected to the equilibrium one by a relaxation type differential equation. It led them to a mechanical wave equation, including a temperature-dependent source term.  Unfortunately, they did not continue in that way later.
  \item Bargmann and Steinmann \cite{BarSte05a, BarSte08} applied the approach of Green and Naghdi \cite{GreenNagh91, GreenNagh92} in order to simulate the coupled wave propagation in solids. Besides thermal expansion, the basic idea behind the Green-Naghdi theory is to introduce 'thermal-displacement', which can be interpreted as a sort of memory effect. Eventually, the previously presented MCV-TE model appears in their model as a special subsystem, but the memory effect has a different formulation.
\end{itemize}
While the existence of non-equilibrium temperature is still an important question \cite{CiaRes16, Restuccia16, JouRes18, JouRes16}, it could not be detected and did not help in the modelling of ballistic heat conduction. Now let us continue with the experimental background.

\subsection{Heat conduction experiments}
The initial success in low-temperature experiments gained increasing attention, leading to a vast number of measurements on various solids and fluids \cite{NarDyn72a, NarDyn75, Acketal66, LaneEtal46, MauHer49}, and these topics are still enjoying a continuous interest \cite{SellCimm15, GuoEtal17, SalJou20}.
Despite the many attempts made in the direction to reliably detect the ballistic effect, there are only a few of them being successful, most of them performed by McNelly et al.~\cite{McN74t}. However, even in these cases, there are many ambiguities, such as the identification of samples, the temperature scale, and so on \cite{KovVan16}. Since there is one particular measurement in the central interest, that is evaluated by several authors; we present that one here.

\begin{figure}[H]
  \includegraphics[width=12cm,height=5cm]{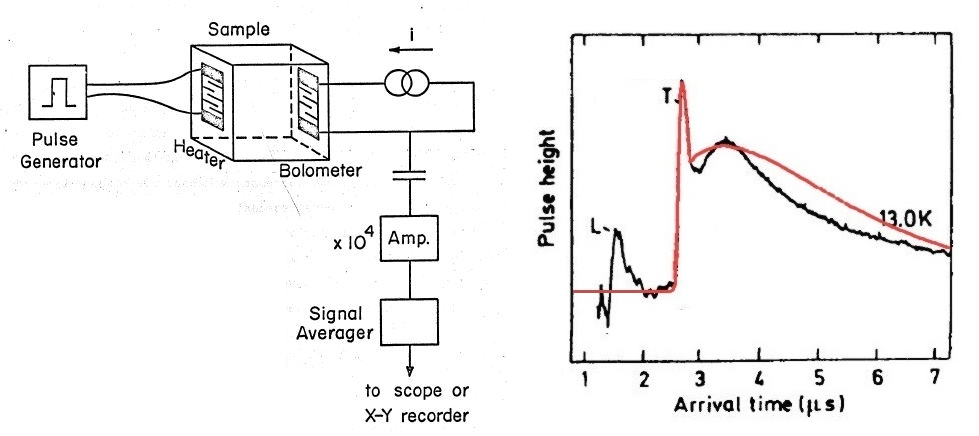}
  \caption{Left side: settings for McNelly's measurements. Right side: the black line presents a typical rear side temperature history for a NaF crystal \cite{McN74t}. Here, L and T are showing the longitudinal and transversal ballistic signals, followed by the second sound. The red line is about demonstrating the modeling capability of the NET-IV model \cite{KovVan18}.}
  \label{fig:exp1}
\end{figure}

The experiment itself is a so-called flash or heat pulse technique. One short heat pulse excites one side of the sample, and the temperature history is recorded on the other side, see Fig.~\ref{fig:exp1} for the arrangement \cite{McN74t}. All temperature curves in the following originate from such measurements. Also, Fig.~\ref{fig:exp1} shows the appearance of all three propagation modes: diffusive, second sound and ballistic signal, each one detected based on their time scales, i.e., the ballistic signal (denoted with L and T) is the fastest one with the speed of sound.

In this particular case, unique crystals are utilized, having extreme purity and exclusively grown for these experiments \cite{Wal63, McN74t}. Regarding their material parameters, we refer to the PhD thesis of McNelly \cite{McN74t}, and also to the paper of Kovács and Ván \cite{KovVan16, KovVan18} which one clarifies some frequent mistakes found in relevant papers about the identification of the crystals. Figure \ref{fig:exp1} demonstrates one example of the temperature history recorded, in which one can observe the three propagation modes altogether. Moreover, that Fig.~\ref{fig:exp1} also shows the earlier results of NET-IV by red, presenting a quantitative reproduction of the experiment with the following parameters:

\begin{table}[H]
  \centering
\begin{tabular}{c|c|c|c|c}
     & \begin{tabular}[c]{@{}c@{}}thermal \\ conductivity\end{tabular} & \begin{tabular}[c]{@{}c@{}}specific \\ heat\end{tabular} & speed of sound & length  \\ \hline
13 K & 10200 W/(mK)                                                    & 1.8 J/(kgK)                                              & 3186 m/s       & 7.9 mm
\end{tabular}
\caption{The known parameters for a NaF crsytal considered here, for $T_0=13$ K reference temperature \cite{McN74t, KovVan18}.}
\label{parametersNAF}
\end{table}

\begin{figure}[H]
  \includegraphics[width=7cm,height=1.5cm]{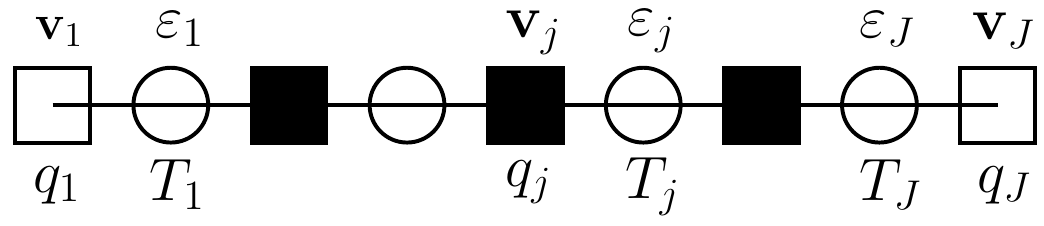}
  \caption{The spatial discretization of the MCV-TE model, having a staggered placement in space. The velocity field $v$ and the temperature field $T$ are shifted with $\Delta x/2$ inside, circles are showing their place.}
  \label{fig:num1}
\end{figure}

Since the temperature is seemingly returning to its initial value, that evaluation required the extension of the energy balance, i.e., the source term now consists of $-h(T-T_0)$ in eq.~\eqref{bceq} with $h$ being the volumetric heat transfer coefficient. This is the effective modelling of a 2D propagation, occurring due to a point-like excitation, and models cooling even with adiabatic boundary conditions. In the RET model, such an effect is implemented with assuming a semi-infinite space. Here, we can solve the equations on a finite space interval, using the length of the crystal.
Consequently, the following system of equations has been solved numerically:
\begin{eqnarray}
\rho c \partial_t T + \partial_x q & = & -h(T-T_0) \nonumber, \\
\tau_q \partial_t q +  q + \lambda \partial_x T + \kappa_{21} \partial_x Q &=& 0  \nonumber , \\
\tau_Q \partial_t Q +  Q - \kappa_{12} \partial_x q&=& 0 . \label{bceq2}
\end{eqnarray}
Its numerical solution required further techniques \cite{RietEtal18} which are worth to present and demonstrate on the MCV-TE model in order to reflect its physical content and the similarities to eq.~\eqref{bceq2}. Let us summarize the MCV-TE model
\begin{align}
  \rho \partial_t v -\partial_x \sigma&=0, \nonumber \\
  \rho \partial_t e+\partial_x q &= \sigma \partial_x v -h(T-T_0), \nonumber \\
  \sigma&=E \varepsilon - E \alpha (T-T_0), \nonumber \\
  e&=cT+\frac{E}{2\rho} \varepsilon^2+\frac{E\alpha}{\rho}(T-T_0) \varepsilon, \nonumber \\
  \tau_q \partial_t q + q &=-\lambda \partial_x T, \nonumber \\
  \partial_t \varepsilon &= \partial_x v. \label{MCVTE}
\end{align}
The key point is the treatment of the boundary conditions. Here, in the MCV-TE model, or also in the ballistic-conductive model of NET-IV, there are multiple field variables coupled to each other. Therefore the boundary conditions are also not independent of each other, which influences how to define them.
For instance, the heat pulse with $t_p$ length prescribes the heat flux on one side such as
\begin{center}
	$q( x=0, t)= \left\{ \begin{array}{cc}
	q_{\textrm{max}}\left(1-\cos\left(2 \pi \cdot \frac{t}{t_p}\right)\right) &
	\textrm{if } 0<  t \leq  t_p,\\
	0 & \textrm{if } t>  t_p,
            \end{array} \right.  $
\end{center}
that is a smooth function convenient to the simulations (and also to the analytical calculations).
The other (rear) side is prescribed to be adiabatic. One question emerges immediately: how to define the other boundaries in a compatible way with this one? The answer remains for future research. Previously, we applied a staggered field discretization for spatial derivatives, avoiding to answer that question. Consequently, only the heat flux remained on the boundary, all the others are shifted with a  half-space step inside \cite{RietEtal18}. It makes that technique to be very similar to finite volume methods, using the local form of the balances instead. Figure \ref{fig:num1} demonstrates how to realize this idea in the case of the MCV-TE model. The difference between the ballistic-conductive and MCV-TE models lies in the structure of the equations: there are two fundamental balances in the latter one. Hence it requires two field quantities to be placed onto the boundaries. In this case, we choose $q$ and $v$, where $v=0$ on both sides describing a fixed surface.

Recently, we found that applying a symplectic discretization for time derivatives leads to an efficient and reliable solution, especially in the facet of energy preservation, such scheme is free from artificial dissipation \cite{FulEtal20, JozsKov20b}. In the present case, it is realized by the simplest method, called symplectic (or semi-implicit) Euler, in which the order of field variable updating is crucial. Thus we begin with the 'volume-like' quantities, that is, calculating the temperature $T$ and the strain $\varepsilon$, then using their updated values, we calculate the heat flux $q$ and the velocity field $v$. The methods are described in detail in \cite{FulEtal20, JozsKov20b}.

\subsubsection{Dimensionless quantities} In the present form, it would be really inconvenient to solve the system \eqref{MCVTE} numerically due to the different time scales, i.e., the mechanical part is significantly faster than the diffusive part for heat conduction. Choosing the following combination of parameters makes the computation faster and more efficient:
\begin{align}
  \hat{v}=\frac{v}{v_0}; \quad v_0=\sqrt{\frac{E}{\rho}}; \quad \hat{x}=\frac{x}{L}; \quad \hat{t}=\frac{v_0 t}{L}; \quad \hat{q}=\frac{q}{q_0}; \quad q_0=q_{\textrm{max}}; \quad \hat{T}=\frac{T}{T_0}
\end{align}
in which $v_0$ is the speed of sound in the one dimensional case, and the quantities with hat denotes their dimensionless attribute. Moreover, we need the dimensionless version of the thermal expansion coefficient: $\hat{\alpha}=\alpha T_0$. Hence the dimensionless system for MCV-TE model reads (without hats)
\begin{align}
  \partial_t v&=\partial_x \varepsilon - \alpha \partial_x T, \nonumber \\
  \partial_t T&=-P_1 \partial_x q - P_2 (T-1) \partial_x v - h_d(T-1), \nonumber \\
  \tau_d \partial_t q + q &= - \lambda_d \partial_x T, \nonumber \\
  \partial_t \varepsilon &= \partial_x v,
\end{align}
where the following dimensionless parameters appeared:
\begin{align}
  P_1=\frac{q_0}{\tilde c v_0}; \quad P_2 = \frac{2E \alpha}{\tilde c}; \quad \tau_d=\frac{\tau_q v_0}{L}; \quad \lambda_d=\frac{\lambda T_0}{q_0 L}; \quad h_d=\frac{h L }{\tilde c v_0}
\end{align}
in which $\tau_d$ and $\lambda_d$ being the dimensionless relaxation time and thermal conductivity, respectively; $a_d$ stands for the dimensionless volumetric heat transfer coefficient. Remarkably, $\tilde c$ denotes an 'effective' specific heat capacity, defined as $\tilde c= \rho c+ E \alpha \varepsilon$, in which the thermal expansion adds a mechanical contribution. Although such contribution could be negligible in general, it has far reaching consequences when nonlinearities are considered as well. Finally, the dimensionless initial condition is $T=1$, and the heat pulse is
\begin{center}
	$q( x=0, t)= \left\{ \begin{array}{cc}
\left(1-\cos\left(2 \pi \cdot \frac{t}{t_p}\right)\right) &
	\textrm{if } 0<  t \leq  t_p,\\
	0 & \textrm{if } t>  t_p,
            \end{array} \right.  $
\end{center}
and the corresponding difference equations are the following:
\begin{align}
  T^{n+1}_j&=-\frac{P_1 \Delta t}{\Delta x}(q^n_{j+1}-q^n_j)-\frac{P_2 (T^n_j-1) \Delta t}{\Delta x}(v^n_{j+1}-v^n_j)-h_d \Delta t(T^n_j-1)+T^n_j, \nonumber \\
  \varepsilon^{n+1}_j&=\frac{\Delta t}{\Delta x} (v^n_{j+1}-v^n_j) + \varepsilon^{n}_j, \nonumber \\
  q^{n+1}_j&=q^n_j\left(1-\frac{\Delta t}{\tau_d} \right) - \frac{\lambda_d \Delta t}{\tau_d \Delta x}(T^{n+1}_j-T^{n+1}_{j-1}), \nonumber \\
  v^{n+1}_j&=\frac{\Delta t}{\Delta x}(\varepsilon^{n+1}_j-\varepsilon^{n+1}_{j-1})-\frac{\alpha \Delta t }{\Delta x}(T^{n+1}_j-T^{n+1}_{j-1})+v^n_j. \label{discMCVTE}
\end{align}

\subsubsection{Stability}
Since the scheme what we utilize here for demonstration is explicit, we must investigate the corresponding stability properties. Following \cite{NumRec07b, KovVan15, FulEtal20, Jury74}, we use the von Neumann method and Jury conditions, which is suitable for linear equations. However, due to the appearance of the nonlinear term $(T-T_0)\partial_xv$, we have to do first the following. Estimating the maximum of the temperature field $T$ is making the nonlinear term to be linear with substituting a constant for $T$ \cite{KovRog20}. Unfortunately, it is an open mathematical question of how to make the proper estimation for $T$. Here, we assume that $C=\max_{(j,n)} (T^n_j-1)$, for all  $x$ and $t$ is 1.
Next, what serves for the starting point, is the discrete version of the dispersion relation, i.e., assuming the solution for the difference equations in the form of a plane wave:
\begin{align}
  u_j^n=\xi^n e^{ikj\Delta x}, \label{vN}
\end{align}
in which $\xi$ is called growth factor, this is the amplitude of the wave and must be bounded from above for stability: $|\xi|\leq 1$. Furthermore, $i$ is the imaginary unit, and $k$ is the wavenumber. After substituting eq.~\eqref{vN} into \eqref{discMCVTE}, we 
obtain the system of linear algebraic equations:
\begin{equation}
\mathbf M\cdot
\left( \begin{array}{llll}
T_0&  \varepsilon_0 &q_0& v_0
\end{array}\right )^T
=0
\end{equation}
where the coefficient matrix is 
{\footnotesize
\begin{equation}\nonumber
\mathbf M=
\left(
\begin{array}{llll}
\xi-1&  0  & \frac{\Delta t}{\Delta x} P_1\left(e^{ik\Delta x}-1\right) &  \frac{\Delta t}{\Delta x} P_2 C \left(e^{ik\Delta x}-1\right)\\ 
0     & \xi-1   &0   & - \frac{\Delta t}{\Delta x} \left(e^{ik\Delta x}-1\right)\\
\frac{\Delta t}{\Delta x} \frac{\lambda_d}{\tau_d}  \xi \left(1-e^{-ik\Delta x}\right)  &0   &\xi-1+\frac{\Delta t}{\Delta x}   &0\\
\frac{\Delta t}{\Delta x} \alpha  \xi \left(1-e^{-i k\Delta x}\right)   &- \frac{\Delta t}{\Delta x} \xi \left(1-e^{-ik\Delta x}\right)    
&0  &\xi -1
\end{array}
\right).
\end{equation}}
Then the characteristic polinomial for $\xi$ can be formally expressed as
\begin{align}
  p(\xi)=a_4\xi^4+a_3\xi^3+a_2\xi^2+a_1\xi+a_0,
\end{align}
where the coefficients are found to be
\begin{align}
  a_4&=1;  \\
  a_3&=\frac{\Delta t \Delta x^2 - 2(\cos(k\Delta x)-1)\Delta t (P_1 \lambda_d + \tau_d + P_2 C \alpha \tau_d)\Big)}
	{\Delta x^2 \tau_d}-4; \nonumber \\
			 a_2&=\frac{4(\cos(k\Delta x)-1)^2 \Delta t^4 P_1 \lambda_d -3 \Delta x^4 (\Delta t - 2 \tau_d)}{\Delta x^4\tau_d}-\nonumber\\
 &-\frac{2(\cos(k\Delta x)-1)\Delta t^2  (\Delta t + \Delta t P_2 C \alpha -2 P_1 \lambda_d -2 \tau_d - 2 P_2 C \alpha \tau_d)}{\Delta x^2\tau_d};\nonumber \\
a_1&=\frac{(3\Delta t - 4 \tau_d)\Delta x^2 +2(\cos(k\Delta x)-1)\Delta t^2  \Big(\Delta t + P_2 C \alpha - P_1 \lambda_d -(1+ P_2 C \alpha \tau_d )\Big)}{\Delta x^2  \tau_d} \nonumber \\
  a_0&=1-\Delta t/\tau_d. \nonumber
\end{align}
The roots of that characteristic polinomial can be restricted into the unit circle on a complex plane, using Jury conditions \cite{Jury74}, i.e.,
\begin{itemize}
  \item $p(\xi=1) \geq 0$;
	  \item $p(\xi=-1) \geq 0$;
  \item $|a_0|\leq |a_4|$;
  \item $|b_0 |>|b_3|$, where
	$b_k= \left | \begin{array}{cc}
a_0 & a_{n-k} \\
a_n & a_k
\end{array} \right | $, with  $n =4 $ and $k=3$; 
 \item $|c_0 |>|c_2|$, where
$c_k= \left | \begin{array}{cc}
b_0 & b_{n-1-k} \\
b_{n-1} & b_k
\end{array} \right | $,  with  $n =4 $ and $k= 2$.
\end{itemize}
Unfortunately, the analytical form of these inequalities does not allow to simplify the expressions reasonably, therefore we decided to check them numerically. However, let us remark that the stability conditions are in agreement with the thermodynamic restrictions, including the relations between the material parameters.
\subsubsection{Demonstrative solution for the MCV-TE model}
\begin{figure}[H]
  \includegraphics[width=12cm,height=6cm]{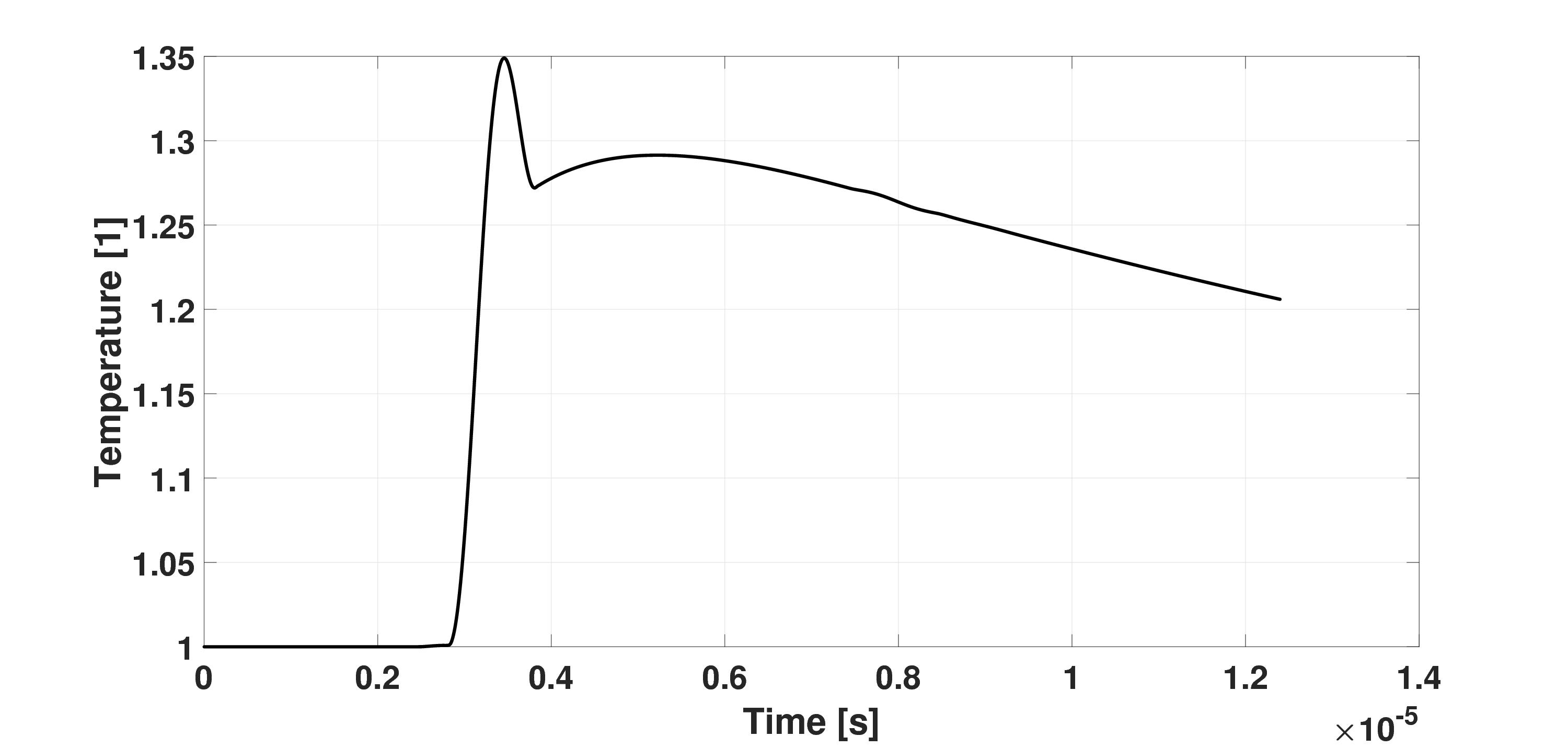}
  \caption{The outcome of a numerical simulation of the MCV-TE model with the NaF data.}
  \label{fig:num2}
\end{figure}

In this particular example, the pulse length was 1 $\mu$s, and the parameters of the Table \ref{parametersNAF} are used. Furthermore, in order to achieve the precise propagation speeds in the simulation, we can fit the mass density $\rho$, the relaxation time $\tau_q$, the volumetric heat transfer coefficient $a$, and the thermal expansion coefficient $\alpha$. Unfortunately, the outcome of McNelly's measurements is without a vertical temperature scale (Fig.~\ref{fig:exp1}), thus the only constraint could be the relative amplitude of the wave signals. We found $\rho=1600$ kg/m$^3$, $a=9 \cdot 10^{-7}$ 1/K, $\tau_q=0.45$ $\mu$s, and $h=0.1626$ W/(mm$^3$K), see Figure \ref{fig:num2} for details.

Apparently, the agreement with the experimental data is remarkable. We emphasize that the mass density is considered to be adjustable in that example, the other parameters are falling the similar order of magnitude what the other authors found earlier in \cite{BarSte05a}. Thus it is still worth to investigate the MCV-TE model in the future.

\subsection{Ballistic-conductive (BC) model from NET-IV}
In that case, the mechanics is missing. Instead, it assumes the existence of an internal variable in the form of a second-order tensor. Therefore we have only one balance equation for the internal energy $e=cT$, and two constitutive equations for $q$ and $Q$. Its numerical treatment can be found in \cite{RietEtal18}, here we use it to validate the analytical solution, presented in the next section.
Since this BC model lies on a different ground, we applied different dimensionless parameters. Following \cite{KovVan15}, specifically for heat pulse experiments, we define
\begin{eqnarray}
\hat{t} =\frac{a t}{L^2} \quad &\textrm{with}& \quad
a=\frac{\lambda}{\rho c};  \quad
\hat{x}=\frac{x}{L};\nonumber \\
\hat{T}=\frac{T-T_{0}}{T_{\textrm{end}}-T_{0}} \quad &\textrm{with}&\quad
T_{\textrm{end}}=T_{0}+\frac{\bar{q}_0 t_p}{\rho c L};  \nonumber \\
\hat{q}=\frac{q}{\bar{q}_0} \quad &\textrm{with}&\quad
\bar{q}_0=\frac{1}{t_p}  \int_{0}^{t_p} q_{0}(t)dt; \nonumber \\
\hat{Q} = \sqrt{- \frac{\kappa_{12}}{\kappa_{21}}} \bar{q}_0 Q, \quad &\textrm{and}& \quad \hat h = h \frac{t_p}{\rho c},
\label{ndvar}
\end{eqnarray}
in which $a$ becomes the thermal diffusivity, $\lambda$ is the thermal conductivity, $T_{end}$ is the equilibrium temperature in the adiabatic situation, $\bar q$ is the mean value of the heat pulse  and $k_{12}$, $k_{21}$ are phenomenological parameters. Furthermore, we need further dimensionless quantities for the material parameters and time scales, correspondingly
\begin{equation}
\hat{\tau}_\Delta =\frac{a t_p}{L^2}; \quad
\hat{\tau}_q 	  = \frac{a \tau_{q}}{L^2}; \quad
\hat{\tau}_Q 	  = \frac{a  \tau_{Q}}{L^2}; \quad
\hat{\kappa} 	  = \frac{\sqrt{-\kappa_{12} \kappa_{21}}}{L},
\end{equation}
where $\hat{\tau}_\Delta$ is the dimensionless pulse length, $\hat{\tau}_q$ and $\hat{\tau}_Q$ are the dimensionless relaxation times and $\hat{\kappa}$ is the dimensionless dissipation parameter.
Now it forms the BC model, exclusively with dimensionless quantities; thus, we neglect the hat notation:
\begin{eqnarray}
{\tau}_{\Delta}\partial_{ t} T +
	\partial_{ x}  q &=& - h T ,\nonumber \\
{\tau}_q \partial_{ t} q +  q +
    {\tau}_{\Delta}\partial_{ x} T +
    {\kappa}\partial_{ x} Q &=& 0
    , \nonumber \\
{\tau}_Q \partial_{ t} Q + Q +
	 \kappa \partial_{ x} q &=& 0 . \label{nd_balcond}
\end{eqnarray}
In Section 3, we focus on the system \eqref{nd_balcond}. Due to the different combination of dimensionless quantities, the initial condition modifies to $T=0$ for the temperature and also homogeneously zero for all other fields. The boundary conditions remain unchanged and given only for $q$.

\section{Analytical solution of the BC model}
While solving the equations \eqref{nd_balcond} numerically with staggering the fields in space, this can be done analytically by eliminating all the other variables what we do not have boundary conditions, i.e., the temperature $T$ and the thermal pressure $Q$ in this case. Let us introduce the upper dot notation for the time derivatives, and from now on, prime denotes the spatial derivative. It gives
\begin{align}
  \tau_q \tau_Q \dddot q + \left(\tau_q+\tau_Q+\frac{h\tau_q \tau_Q}{\tau_{\Delta}}\right)\ddot q+ \left(1+\frac{h(\tau_q + \tau_Q)}{\tau_\Delta} \right) \dot q + \frac{h}{\tau_\Delta} q =& \nonumber\\
  =\left(\frac{h(\kappa^2+\tau_Q)-\tau_Q h - \tau_\Delta}{\tau_\Delta} \right) q'' +(\kappa^2+\tau_Q) \dot q''. \label{bcqh}
\end{align}
Equation \eqref{bcqh} is a reaction-diffusion type one since the source $hT$ in the energy balance yields a zeroth order term in $q$. That term can be eliminated simply using
\begin{align}
  q=e^{-\frac{h}{\tau_\Delta}t}\tilde q
\end{align}
transformation, and it results in
\begin{align}
  \tau_q \tau_Q \dddot{\tilde q} + \left(\tau_q+\tau_Q-2\frac{h \tau_q \tau_Q}{\tau_\Delta}   \right) \ddot{\tilde q} + \left(1 - \frac{h (\tau_q + \tau_Q)}{\tau_\Delta} + \frac{h^2 \tau_q \tau_Q}{\tau_\Delta^2} \right) \dot{\tilde q} =& \nonumber \\
  =\left( -\frac{\tau_Q h - \tau_\Delta}{\tau_\Delta} \right) \tilde q'' + (\kappa^2+\tau_Q) \dot{\tilde q}''. \label{bcqh2}
\end{align}
Now, let us simplify this equation with $h=0$ because it does not give any further insight, and the solution still able to validate the numerical code, which is our primary purpose. Thus eq.~\eqref{bcqh2} reduces to
\begin{align}
  \tau_q \tau_Q \dddot{\tilde q} + \left(\tau_q+\tau_Q \right) \ddot{\tilde q} + \dot{\tilde q} = \tilde q'' + (\kappa^2+\tau_Q) \dot{\tilde q}'', \label{bcqh3}
\end{align}
in which we further simplify our notations by introducing the parameters $a=\tau_Q\tau_q$, $b=\tau_Q+\tau_q$, $c=\tau_Q+\kappa^2$, and hence the tilde becomes unnecessary ($q=\tilde q$):
\begin{equation}
a \dddot{q} + b \ddot{q} + \dot{q}= q'' + \dot{q}''.
\label{eq:4}
\end{equation}
Summarizing the initial and boundary conditions:
\begin{equation}
    q(x=0,t)=q_0(t)=
    \begin{cases}
      1-\cos(2\pi \frac{t}{\tau_{\Delta}}), &  t \leq \tau_{\Delta} \\
      0, & t>\tau_{\Delta}
    \end{cases}
\label{eq:5}
  \end{equation}
and $q(x,t=0)=\dot{q}(x,t=0)=\ddot{q}(x,t=0)=T(x,t=0)=Q(x,t=0)=q(x=1,t)=0$. There are some important consequences of such decision. From now on, it is no longer possible to implement boundary conditions with temperature, and the classical Neumann ('second-type') boundary becomes a first-type one, therefore it is easier to handle a time-dependent boundary within this setting.
On the other hand, it is entirely arbitrary which field variable to use, and it is also possible to use either $T$ or $Q$, depending on the physical situation. However, having one generalized constitutive relation between $T$ and $q$ (or $Q$, accordingly) makes it impossible to classically interpret the 'first-type' and 'second-type' boundary conditions. Moreover, it is still an open question of what happens with the Robin ('third-type') boundary in such a model, and whether the constitutive equation affects their definition or not.

These aspects have a crucial role in solving generalized equations. Otherwise, one might encounter false solutions, and the content seemingly violates basic mathematical and physical
principles such as the maximum principle and second law of thermodynamics. For instance, not surprisingly, using unphysical settings for the MCV equation, it easily leads to unphysical temperature evolution, contradicting the second law by achieving negative values \cite{HuCao09}. Negative temperature behavior is also observable for the Guyer-Krumhansl (GK) equation (having $\tau_Q=0$ in eq.~\eqref{eq:4}) \cite{Zhukov16, Zhu16a, Zhu16b, ZhuSri17}. We believe that the above-mentioned problems could have an impact, even if it is invisible at this moment.
We note that other analytical solutions do not reflect the unphysical behaviour of both the MCV and GK equations \cite{Kov18gk}.

The complete solution of eq.~\eqref{eq:4} is splitted into two parts. The first one lasts until the end of the heat pulse $\tau_\Delta$, while the next one is having a constant boundary, independelty of the time.
\subsection{Section I ($0<t<t_\Delta$)}
Starting with the first section, let us assume that we can write the heat flux in the form $q=v+w$, in which $w$ serves to separate the time-dependent inhomogeneous boundary from $v$ which becomes the homogeneous solution of eq.~\eqref{eq:4}, and
\begin{equation}
q=v+w; \quad  w = \Big( 1-\frac{x}{L} \Big) q_0(t).
\label{eq:6}
\end{equation}
Substituting eq.~\eqref{eq:6} into \eqref{eq:4}, and executing the derivatives, we obtain the following inhomogeneous partial differential equation:
\begin{equation}
a \dddot{v}+b\ddot{v}+\dot{v}=v''+c\dot{v}''-f(x,t),
\label{eq:7}
\end{equation}
where $f(x,t)=a\dddot{w}+b\ddot{w}+\dot{w}$. Using the splitting $v=q-w$, the initial conditions for $v$ are not completely equal to zero, that is,
\begin{align}
  v(x,t=0)=\dot{v}(x,t=0)=0, \quad \ddot{v}(x,t=0) = \frac{4\pi^2 x}{\tau_{\Delta}^2L}.
\end{align}
Applying the classical procedure of separation of variables, we seek the solution in the form $v(x,t)=\phi(t)X(x)$, in which $\phi(t)$ describes the time evolution, and $X(x)$ stands for the spatial behavior. Furthermore, following \cite{Kov18gk}, we assume that the inhomogeneity $f(x,t)$ can be represented in the space spanned by the eigenfunctions of Laplacian operator. Thus we determine the eigenspace for the homogeneous part ($f(x,t)=0$),
\begin{equation}
\frac{a\dddot{\phi}+b\ddot{\phi}+\dot{\phi}}{\phi+c\dot{\phi}}=\frac{X''}{X},
\label{eq:8}
\end{equation}
where $X''/X = \beta$ yields the eigenvalues and the eigenfunctions, with the homogeneous boundaries $X(0)=X(L)=0$:
\begin{equation}
X_n(x)=\sin(\sqrt{\beta_n}x),
\label{eq:9}
\end{equation}
with $\beta_n = \Big( \frac{n\pi}{L}\Big)^2$ being the eigenvalues. Then, exploiting the linearity of eq.~\eqref{eq:4}, $v(x,t)$ reads
\begin{equation}
v(x,t)=\sum_{n=1}^{\infty} \phi_n(t)\sin\Big(\frac{n\pi}{L}x\Big).
\label{eq:10}
\end{equation}
Now, we can substitute eq.~\eqref{eq:10} into \eqref{eq:7}:
\begin{equation}
\sum_{n=1}^{\infty}\Big[ a\dddot{\phi}_n + b \ddot{\phi}_n + (1+c\beta_n)\dot{\phi}_n  + \beta_n \phi_n \Big]\sin\Big( n\pi \frac{x}{L}\Big) = -f(x,t),
\label{eq:11}
\end{equation}
which equation requires the Fourier sin-series expansion of $f(x,t)$, i.e., $f(x,t)=\sum_{n=1}^{\infty}f_n(t)\sin\Big( n\pi\frac{x}{L}\Big)$, and $f_n(t)$ is expressed using eq.~\eqref{eq:5} and \eqref{eq:6}:
\begin{eqnarray}
\label{eq:12}
& {\displaystyle f_n(t)=\frac{2}{L}\Bigg[\Big(\frac{2\pi}{\tau_{\Delta}} -a \frac{8\pi^3}{\tau_{\Delta}^3}\Big) \sin\Big( 2\pi\frac{t}{\tau_{\Delta}}\Big)  + b\frac{4\pi^2}{\tau_{\Delta}^2}\cos\Big( 2\pi\frac{t}{\tau_{\Delta}}\Big) \Bigg] \times }\\ \nonumber
& {\displaystyle \times \int_{0}^{L} \Big( 1-\frac{x}{L}\Big) \sin\Big( \frac{n\pi}{L}x\Big) dx. }
\end{eqnarray}
After integration and simplification, we obtain the n-th term as follows:
\begin{equation}
f_n(t) = \frac{2}{n\pi}\Bigg[ \Big( \frac{2\pi}{\tau_{\Delta}}-\frac{8a\pi^3}{\tau_{\Delta}^3}\Big) \sin\Big( 2\pi \frac{t}{\tau_{\Delta}} \Big) + \frac{4b\pi^2}{\tau_{\Delta}^2}\cos\Big( 2\pi \frac{t}{\tau_{\Delta}}\Big) \Bigg].
\label{eq:12}
\end{equation}
Now, using eq.~\eqref{eq:12}, eq.~\eqref{eq:11} shall lead to the solution of the time evolution for $v$, i.e., $\phi_n(t)$ is prescribed as
\begin{equation}
a \dddot{\phi}_n +b\ddot{\phi}_n+(1+c\beta_n)\dot{\phi}_n+\beta_n\phi_n=-f_n(t),
\label{eq:13}
\end{equation}
and its general solution is
\begin{equation}
\phi_n = Y_{1_n} e^{x_1 t} + Y_{2_n} e^{x_2 t} + Y_{3_n} e^{x_3 t} + A \sin(g t) + B \cos(g t)
\label{eq:14}
\end{equation}
where $x_{1,2,3}$ are the roots of the characteristic polinomial $ax^3+bx^2+(1+c\beta)x+\beta=0$,  $g=2\pi / \tau_{\Delta}$, and $A, \ B$ are the parameters of the particular solution $\phi_p$.
Now substituting $\phi_p=A\sin(gt)+B\cos(gt)$ into eq.~\eqref{eq:13},
\begin{equation}
a \dddot{\phi}_p + b \ddot{\phi}_p + (1+c\beta_n)\dot{\phi}_p+\beta_n\phi_p = K_1 \sin(gt) + K_2\cos(gt),
\label{eq:15}
\end{equation}
with
\begin{align}
K_1 = -\frac{2}{n \pi}\Bigg(   \frac{2\pi}{\tau_{\Delta}} - \frac{8a\pi^3}{\tau_{\Delta}^3} \Bigg); \quad K_2 =  - \frac{2}{n\pi}\frac{4b\pi^2}{\tau_{\Delta}^2},
\end{align}
as well as
\begin{align}
-aAg^3-bBg^2+A(1+\beta_nc)g+\beta_n B&=K_2, \nonumber \\
aBg^3-bAg^2-Bg(1+\beta_nc)+\beta_n A&=K_1, \label{eq:17}
\end{align}
which system gives the expressions for $A$ and $B$.
In order to determine the coefficients $Y_{1_n,2_n,3_n}$, we must take the initial conditions into account:
\begin{align}
\phi(0) &= Y_{1_n}+Y_{2_n}+Y_{3_n}+B=0, \nonumber \\
\dot{\phi}(0) &= x_1Y_{1_n}+x_2Y_{2_n}+x_3Y_{3_n}+Ag=0, \nonumber \\
\ddot{\phi}(0) &= x_1^2Y_{1_n}+x_2^2Y_{2_n}+x_3^2Y_{3_n}-Bg^2=\frac{4\pi^2 x}{\tau_{\Delta}^2L}.
\end{align}
Formally, we obtained the solution of the heat flux $q$ for the first session that lasts until $\tau_\Delta$,
\begin{eqnarray}
\label{eq:21}
\nonumber
&{\displaystyle q_I(x,t) = \sum_{n=1}^{\infty}\Big( Y_{1_n} e^{x_1 t} + Y_{2_n} e^{x_2 t} + Y_{3_n} e^{x_3 t} + A \sin(g t) + B \cos(g t) \Big) \sin(\frac{n\pi}{L}x) +  } \\
&{\displaystyle + \Big(1-\frac{x}{L} \Big) q_0(t) }.
\end{eqnarray}
We can use this expression to reconstruct the temperature field, utilizing the balance of internal energy:
\begin{eqnarray}
\label{eq:22}
\nonumber
&{\displaystyle T_I(x,t) =\frac{t}{L\tau_{\Delta}} - \frac{1}{2\pi L}\sin(gt) -} \\
&{\displaystyle   - \sum_{n=1}^{\infty}\frac{n\pi \cos(\frac{n\pi}{L}x)}{L \tau_{\Delta}}  \Bigg(  \Bigg[ \sum_{j=1}^3 \frac{e^{x_jt}-1}{x_j}Y_{jn} \Bigg]  + \frac{B_n}{g}\sin(gt)+\frac{A_n-A_n\cos(gt)}{g}\Bigg)}.
\end{eqnarray}
\subsection{Section II ($t>\tau_\Delta$)}
Now turning our attention to the second session of the solution (in which $t\geq \tau_\Delta$), we introduce a new variable for time such as $\tilde{t}=t-\tau_{\Delta}$ for convenience. Thus the initial conditions become
\begin{align}
q_{II}(x,\tilde{t}=0) &= q_I(x,t=\tau_{\Delta}), \nonumber \\
\dot{q}_{II}(x,\tilde{t}=0) &= \dot{q}_I(x,t=\tau_{\Delta}), \nonumber \\
\ddot{q}_{II}(x,\tilde{t}=0) &= \ddot{q}_I(x,t=\tau_{\Delta}).
\label{eq:23}
\end{align}
Since the inhomogeneity of $f(x,t)$ disappears as $w\equiv0$, we can immediately begin with
\begin{equation}
a \dddot{\gamma}_n +b\ddot{\gamma}_n+(1+c\beta_n)\dot{\gamma}_n+\beta_n\gamma_n=0.
\label{eq:24}
\end{equation}
The roots of the characteristic polinomial $x_{1,2,3}$ remain unchanged, and the general solution reads
\begin{equation}
\gamma_n(t)=C_{1_n}e^{x_1t}+C_{2_n}e^{x_2t}+C_{3_n}e^{x_3t}.
\label{eq:25}
\end{equation}
in which the unknown $C_{1,2,3}$ coefficients are determined using the initial conditions \eqref{eq:23}:
\begin{align}
C_{1_n}+C_{2_n}+C_{3_n}=\phi_n(t=\tau_{\Delta}), \nonumber \\
x_1C_{1_n}+x_2C_{2_n}+x_3C_{3_n}=\dot{\phi}_n(t=\tau_{\Delta}), \nonumber \\
x_1^2C_{1_n}+x_2^2C_{2_n}+x_3^2C_{3_n}=\ddot{\phi}_n(t=\tau_{\Delta}).
\label{eq:26}
\end{align}
Again, the temperature distribution is recovered using the balance equation.
\begin{equation}
\label{eq:29}
 T_{II}(x,\tilde{t}) = - \sum_{n=1}^{\infty}\frac{n\pi \cos(\frac{n\pi}{L}x)}{L \tau_{\Delta}}    \Bigg[ \sum_{j=1}^3 \frac{e^{x_j\tilde{t}}-1}{x_j}C_{j_n} \Bigg]  + T_{I}(x,t=\tau_{\Delta} ) ,
\end{equation}
where $\tilde{t}=t-\tau_{\Delta}$, and $T_{I}(x,t=\tau_{\Delta})$ is defined by eq.~\eqref{eq:22}.

\subsubsection{Calculating $Q(x,t)$}
The heat current density $Q(x,t)$ can be calculated from eq.~\eqref{nd_balcond} using the previously determined $T(x,t)$ and $q(x,t)$ functions. The solution again is divided into two parts, namely $Q_{I}(x,t)$ and $Q_{II}(x,t)$, where the $T_{I,II}(x,t)$ and $q_{I,II}(x,t)$ functions are used accordingly. The necessary derivatives needed for $Q(x,t)$ are the following:

\begin{eqnarray}
\nonumber
&{\displaystyle \partial_t q_{I}(x,\hat{t}) = \sum_{n=1}^{\infty} \Bigg[ \sum_{k=1}^{3}Y_{kn}x_{k}e^{x_k t} + A_ng\cos(gt) - B_ng\sin(gt)         \Bigg]\sin\Big( \frac{n\pi}{L}x\Big) + } \\
&{\displaystyle+ \Big( 1-\frac{x}{L} \Big) g \sin(gt)},
\end{eqnarray}
\begin{eqnarray}
\nonumber
&{\displaystyle \partial_x T_{I}(x,\hat{t}) = \sum_{n=1}^{\infty} \frac{(n\pi)^2\sin\Big( \frac{n\pi}{L}x \Big)}{L^2\tau_{\Delta}}\Bigg[ \sum_{k=1}^3 \frac{e^{x_k t}-1}{x_k}Y_{kn}   +  \frac{B_n}{g}\sin(gt) +} \\
&{\displaystyle+ \frac{A_n-A_n\cos(gt)}{g} \Bigg]},
\end{eqnarray}

\begin{equation}
\partial_t q_{II}(x,\hat{t}) = \sum_{n=1}^{\infty} \sum_{k=1}^{3}C_{kn}x_{k}e^{x_k\hat{t}}\sin\Big( \frac{n\pi}{L}x\Big),
\end{equation}
\begin{eqnarray}
\nonumber
&{\displaystyle \partial_x T_{II}(x,\hat{t}) = \sum_{n=1}^{\infty} \frac{(n\pi)^2\sin\Big( \frac{n\pi}{L}x \Big)}{L^2\tau_{\Delta}}\Bigg[ \sum_{k=1}^3 \frac{e^{x_k \hat{t}}-1}{x_k}C_{kn} + \sum_{k=1}^3 \frac{e^{x_k \tau_{\Delta}}-1}{x_k}Y_{kn}  +} \\
&{\displaystyle + \frac{B_n}{g}\sin(g\tau_{\Delta}) + \frac{A_n-A_n\cos(g\tau_{\Delta})}{g} \Bigg]},
\end{eqnarray}
Putting everything together the heat flux density for the first time interval is:
\begin{equation}
Q_{I}(x,t)=\frac{1}{\kappa} \sum_{n=1}^{\infty} \frac{L\cos\Big( \frac{n\pi}{L}x\Big)}{n\pi}\mathcal{F}(A_n,B_n,t,x_k) -\Big( x-\frac{x^2}{2L}\Big)\Big( 1-\cos(gt)+\tau_qg\sin(gt)\Big),
\end{equation}
where
\begin{eqnarray}
\nonumber
&{\displaystyle \mathcal{F}(A_n,B_n,t,x_k) =  \ A_n\sin(gt)+B_n\cos(gt)+\sum_{k=1}^3 Y_{kn}e^{x_kt} + \tau_q A_ng\cos(gt)-\tau_qB_ng\sin(gt) + } \\
&{\displaystyle + \tau_q\sum_{k=1}^3Y_{kn}x_k e^{x_kt} + \frac{n^2\pi^2 B_n}{L^2 g}\sin(gt) + \frac{n^2\pi^2}{L^2}\frac{A_n-A_n \cos(gt)}{g} +\frac{n^2\pi^2}{L^2}\sum_{k=1}^3 \frac{e^{x_kt}-1}{x_k}Y_{kn} }.
\end{eqnarray}
Similiarly, the heat flux density for the second interval will be:
\begin{equation}
Q_{II}(x,t) = \frac{1}{\kappa}\sum_{n=1}^{\infty}\frac{L\cos\Big( \frac{n\pi}{L}x\Big)}{n\pi}\mathcal{G}(A_n,B_n,x_k,\hat{t},\tau_{\Delta}),
\end{equation}
in which
\begin{eqnarray}
\nonumber
&{\displaystyle \mathcal{G}(A_n,B_n,x_k,\hat{t},\tau_{\Delta}) = \sum_{k=1}^{3}C_{kn}e^{x_k\hat{t}} + \tau_q\sum_{k=1}^3 C_{kn}x_ke^{x_k\hat{t}}+\frac{n^2\pi^2}{L^2}\Bigg[\sum_{k=1}^3 \frac{e^{x_k\hat{t}}-1}{x_k}C_{kn} +  } \\
&{\displaystyle+ \sum_{k=1}^3 \frac{e^{x_k\tau_{\Delta}}-1}{x_k}Y_{kn} + \frac{B_n}{g}\sin(g\tau_{\Delta})+ \frac{A_n-A_n\cos(g\tau_{\Delta})}{g}   \Bigg] }.
\end{eqnarray}

\section{Validation and conclusions}
\begin{figure}
  \includegraphics[width=12cm,height=6cm]{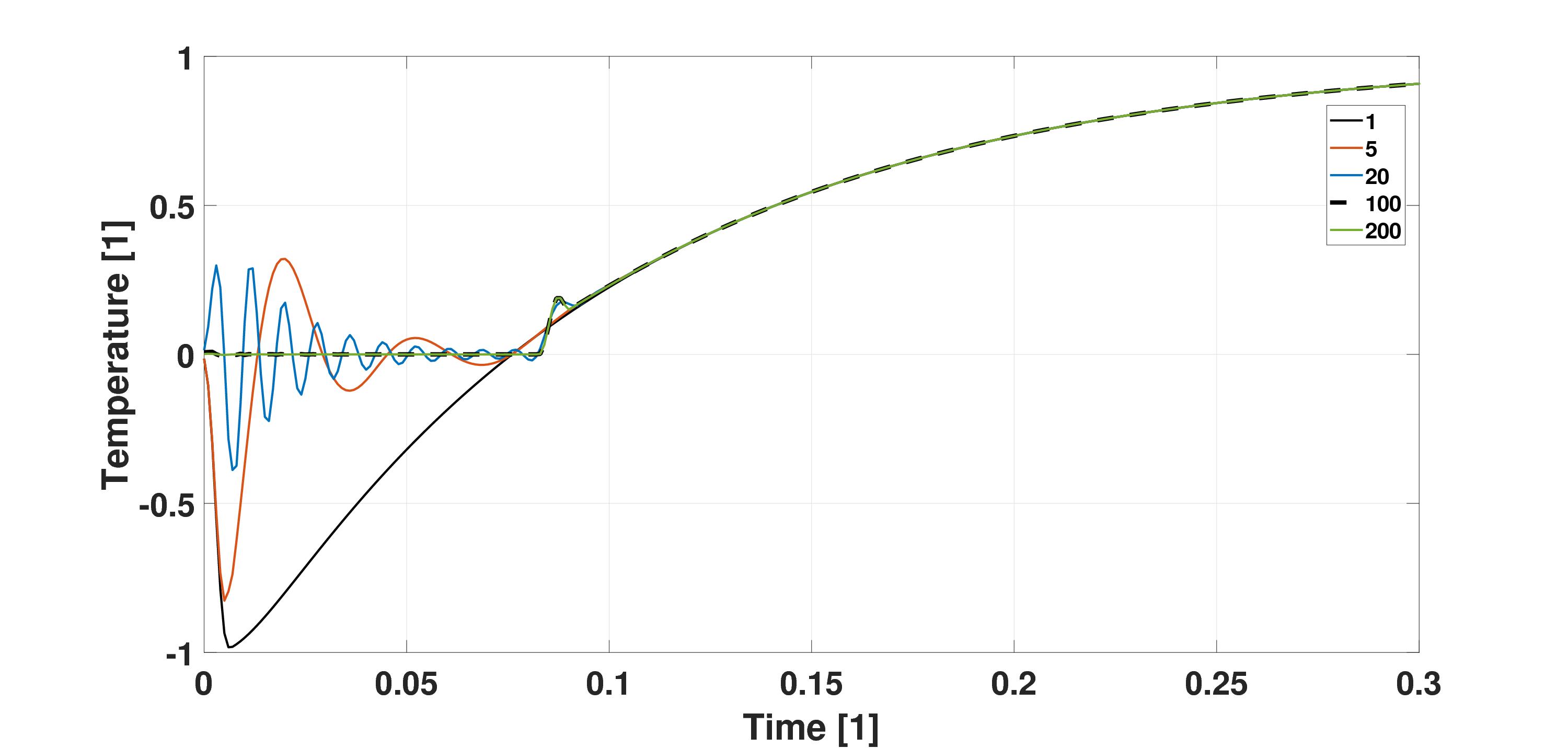}
  \caption{Demonstrating the convergence of the analytical solution with depicting $1, 5, 20, 100$ and $200$ terms. The convergence analysis used the following parameters: $\tau_\Delta=0.0076; \tau_q=0.0113; \tau_Q=0.007; \kappa=0.0663$.}
  \label{fig:anal1}
\end{figure}

Previously we found the temperature field $T$ in the form of an infinite series. First, let us present the convergence of the series by utilizing more and more terms in the series, see Fig.~\ref{fig:anal1} demonstrates this result. It is apparent that there is no significant difference between the last two cases with $100$ and $200$ terms, only in the very beginning of history.
The well-known classical approximation for the analytical solution of the Fourier heat equation appears similarly for the BC model as well; however, with different Fourier number. That is, in the Fourier case, merely the first term of the infinite series can be used for time intervals higher than $0.2$. In the generalized case, it holds, too, even with smaller Fourier number, around $0.1$.

\begin{figure}
  \includegraphics[width=12cm,height=6cm]{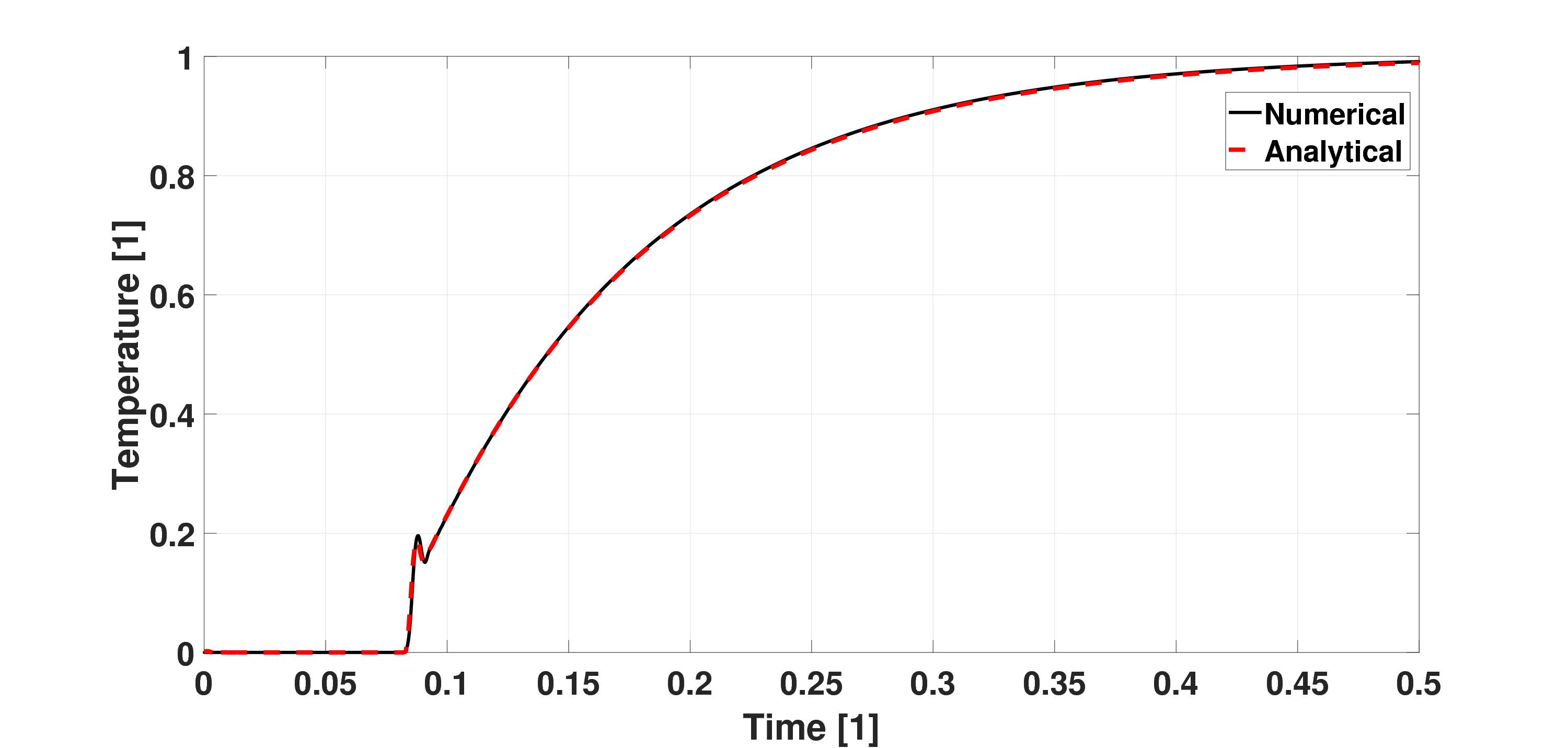}
  \caption{Comparing the analytical solution to the numerical one using $200$ terms, and $\tau_\Delta=0.0076; \tau_q=0.0113; \tau_Q=0.007; \kappa=0.0663$.}
  \label{fig:anal2}
\end{figure}

In order to validate our solution methods, we decided to choose a particular set of parameters:
$\tau_\Delta=0.0076; \tau_q=0.0113; \tau_Q=0.007; \kappa=0.0663$, which are corresponding to the NaF evaluation \cite{KovVan18}, without the source term for cooling. In Fig.~\ref{fig:anal2}, we can observe the appearance of the ballistic signal, even if it is minimal but real. Using $200$ terms can reliably resolve that phenomenon, in the agreement of the numerical solution.
Moreover, we note here that the numerical method is proved to be stable and consistent; therefore, convergence is satisfied in the light of the Lax principle \cite{RietEtal18}.

\begin{figure}[H]
  \includegraphics[width=12cm,height=6cm]{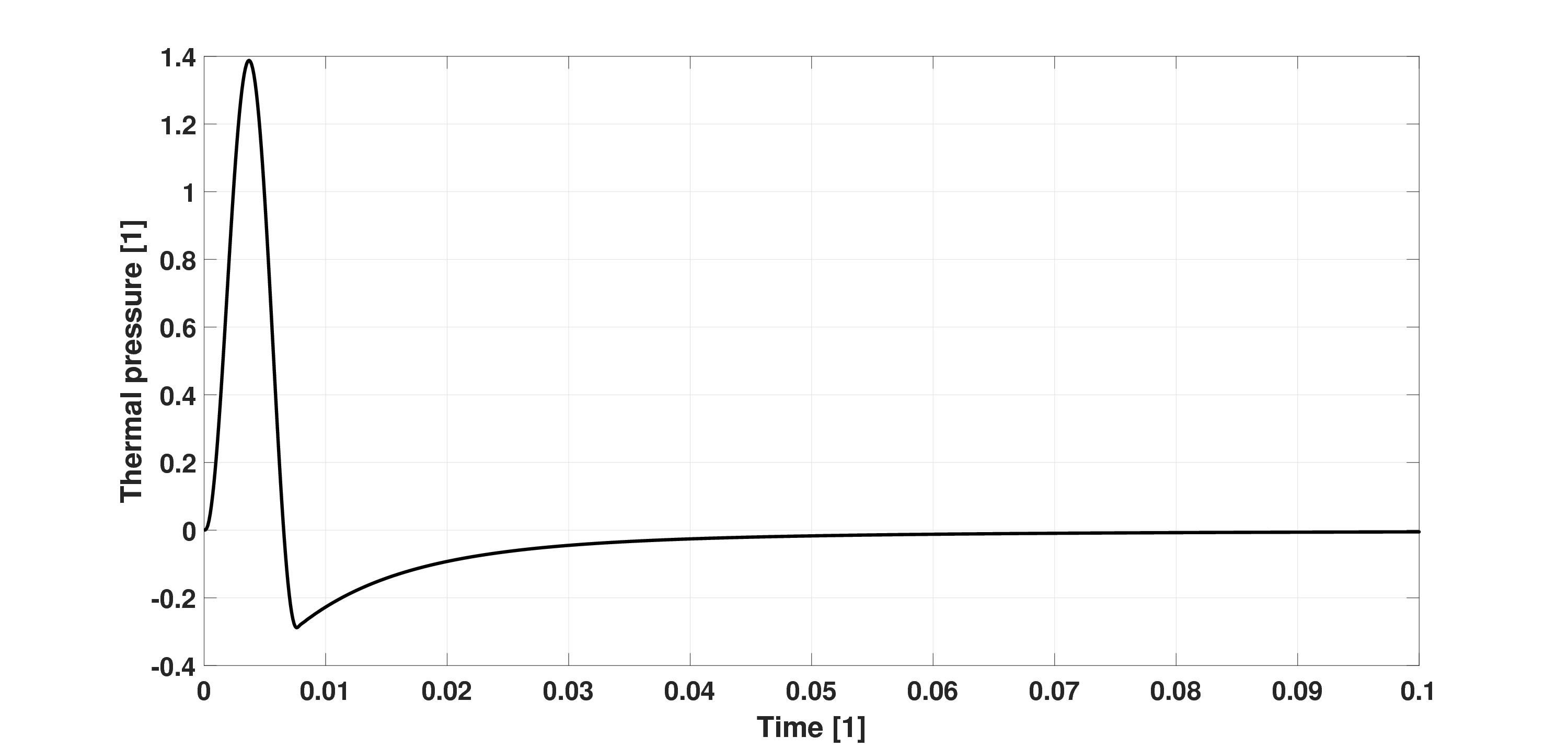}
  \caption{The behaviour of the thermal pressure $Q$ on the front side, using the same parameter set.}
  \label{fig:anal3}
\end{figure}

\begin{figure}[H]
  \includegraphics[width=12cm,height=6cm]{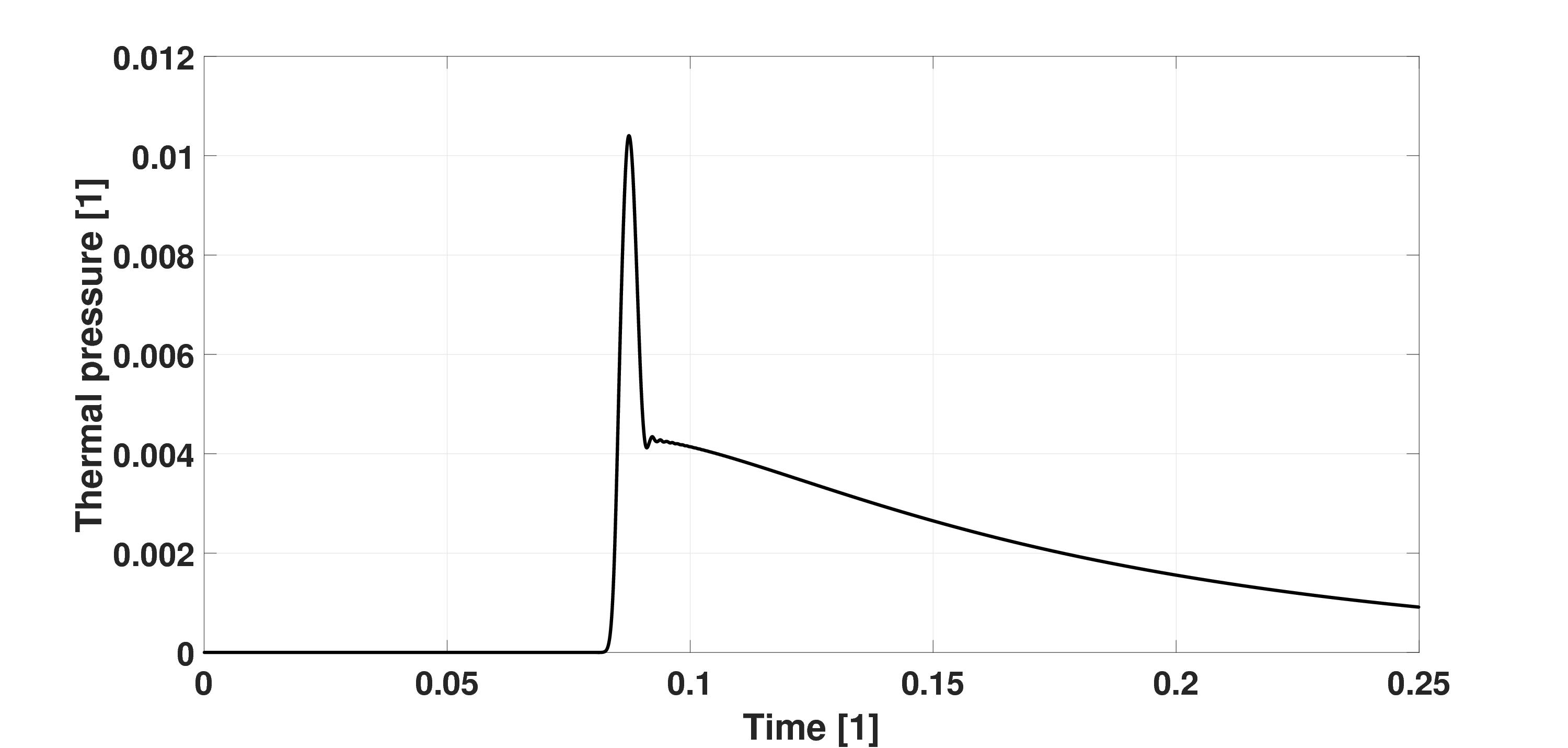}
  \caption{The behaviour of the thermal pressure $Q$ on the rear side, using the same parameter set.}
  \label{fig:anal4}
\end{figure}

Figures \ref{fig:anal3} and \ref{fig:anal4} are showing the behaviour of the thermal pressure $Q$ on the boundaries, accordingly. Although a similar excitation appears on the front side, its implementation as a boundary condition is not trivial. On one hand, it returns to its initial value after the excitation. On the other hand, it is not a simple relaxation process since it requires the gradient of the heat flux, which is unknown before the solution. Regarding the rear side, it is more similar to the temperature history from the NaF experiments, but it is for the thermal pressure $Q$. Its maximum occurs when the ballistic wave enters the rear side and scatters. The following, smaller signal is for the second sound. It seems to be impossible at this moment how to use that data as a boundary condition before the evaluation of experiments.

This is the primary reason why we insist on eliminating the temperature and the pressure, and thus using only the heat flux $q$, at least in such situation when the temperature on the boundary does not play a role. We find it important to mention the properties of the NaF crystals. According to the thermal conductivity measurements of McNelly \cite{McN74t}, these crystals are anisotropic, hence showing direction-dependent behaviour. In such a situation, the model becomes significantly more complicated; thus, we neglected that fact. Also, we neglected the source term as well, since its contribution would not add to the message of this paper.

Notwithstanding, there are numerous ways how to proceed with the research in the future. For instance, the crystals are not 'simply' anisotropic but showing some complex mechanical behaviour as well in low-temperature conditions. In this regard, we refer to the work of Mezhov-Deglin et al.~\cite{Mezhov79, Mezhov80, MezMuk11} for solids, and the work of Sciacca et al.~\cite{SciSellJou14} for fluids.

As an outlook for future research, we repeat that the structure of the BC model is indeed the same as for the phonon hydrodynamic equations with $M=2$. Consequently, the analytic solution method is inherited in these cases. Furthermore, it is also relevant for rarefied gas models \cite{Arietal15, RugSug15, Kov18rg, KovEtal18rg, LebCloo89, CarMorr72a, CarMorr72b, KovJouRog19} in which the same coupling happens. Moreover, it could be useful for both shock structure analysis \cite{MadSim19, Madjar15, SimEtal15} and rheological models \cite{SzucsFul18b, SzucsFul19}. In these cases, $Q$ is explicitly said the be the pressure, and $\nabla q$ is directly contributing to the mechanical processes. We aim to investigate both the analytical and numerical solution methods for these models in the future.

Finally, we aim to keep in our mind the comparison with other approaches beyond EIT, and RET. For instance, the framework of GENERIC \cite{GrmOtt97, OttGrm97, Ott05b, GrmKliPav19, Grmela2018b} is constructive either on the ground of the structure of the equations or about the solution methods, especially on the numerical ones with symplectic schemes such as in the papers \cite{Romero10I, Romero10II, PortEtal17, ShanOtt20}.

\section{Acknowledgements}
The authors thank Mátyás Szücs for valuable discussions.

The research reported in this paper was supported by the grants of National Research, Development and Innovation Office – NKFIH 130378, and by FIEK-16-1-2016-0007. The work was supported by the National Research, Development and Innovation Fund (TUDFO/51757/2019-ITM), Thematic Excellence Program.

The authors acknowledge the financial support of the Italian Gruppo Nazionale per la Fisica Matematica (GNFM-INdAM).

\bibliographystyle{unsrt}

\end{document}